\documentclass{elsart3p}

\usepackage{amssymb,graphicx}

\begin{document}

\begin{frontmatter}

\title{\vspace{-4mm} About the relation between the quasiparticle Green's function in cuprates obtained from ARPES data and the magnetic susceptibility}

\author[a]{D.\,S.~Inosov}
\author[a]{S.\,V.~Borisenko}
\author[b]{I.~Eremin}
\author[a,c]{A.\,A.~Kordyuk}
\author[a]{V.\,B.~Zabolotnyy}
\author[a]{J.~Geck}
\author[a]{A.~Koitzsch}
\author[a]{J.~Fink}
\author[a]{M.~Knupfer}
\author[a]{B.~B\"uchner}

\address[a]{Institute for Solid State Research, IFW Dresden, P.O.Box 270116, D-01171 Dresden, Germany}
\address[b]{Max Planck Institute for the Physics of Complex Systems, D-01187 Dresden, Germany}
\address[c]{Institute of Metal Physics of National Academy of Sciences of Ukraine, 03142 Kyiv, Ukraine\vspace{-3mm}}

\begin{abstract}

Angle resolved photoemission spectroscopy (ARPES) provides a detailed view of the renormalized band structure in
cuprates and, consequently, is a key to the self-energy and the quasiparticle Green's function. Such information gives
a clue to the comparison of ARPES with scanning tunneling microscopy, inelastic neutron scattering (INS), and Raman
scattering data. Here we touch on a potential possibility of such a comparison with the dynamical magnetic
susceptibility measured in INS experiments. Calculations based on the experimentally measured quasiparticle
self-energies in cuprates lead to the estimated magnetic susceptibility response with many-body effects taken into
account. \vspace{-2mm}

\end{abstract}

\begin{keyword}

cuprate superconductors \sep dynamical magnetic susceptibility \sep ARPES \sep inelastic neutron scattering \sep
itinerant magnetism

\PACS 74.72.-h \sep 74.72.Hs \sep 74.25.Ha \sep 74.20.-z \sep 79.60.-i

\end{keyword}
\end{frontmatter}


The normal-state Lindhard response function (polarization operator) is related to the quasiparticle Green's function
via a simple autocorrelation formula \cite{AbanovChubukov99}:

\vspace{-2mm}
\begin{equation}
\hspace{-0.3mm}\chi_0(\textbf{Q},\Omega)\hspace{-0.5mm}\propto\hspace{-0.4mm}
-2i\hspace{-0.9mm}\int{\hspace{-0.8mm}G(\textbf{k},\omega)\hspace{-0.1mm}G(\textbf{k}+\textbf{Q},\omega+\Omega)\hspace{0.2mm}d^2k
\hspace{0.2mm}d\omega ~} \label{eq1}
\end{equation}
\vspace{-2mm}

\noindent Apart from the bare band structure, equation (\ref{eq1}) also holds for the renormalized Green's function.

Knowing the Lindhard response function $\chi_0$ (which is also known as the bare spin susceptibility), one can obtain
from RPA the dynamic spin susceptibility $\chi$ \cite{LiuZhaLevin95}, the imaginary part of which is directly
proportional to the measured INS intensity \cite{SchnyderManske06}:

\vspace{-1mm}
\begin{equation}
\chi(\textbf{Q},\Omega) = {\chi_0(\textbf{Q},\Omega)}/{[1+J_Q\,\chi_0(\textbf{Q},\Omega)]} \label{eq2}
\end{equation}
\vspace{-1mm}

The coefficient $J_Q$ in the denominator of (\ref{eq2}) describes the effective four-point vertex (Hubbard interaction
or superexchange), which can be refined to the required degree of accuracy, although for many applications its
$k$-dependence is neglected \cite{Norman00}. In our calculations we assume $J_Q=J_0\,[\cos{Q_xa}+\cos{Q_ya}]$.

\begin{figure*}
\vspace{-6mm}
\includegraphics[width=\textwidth]{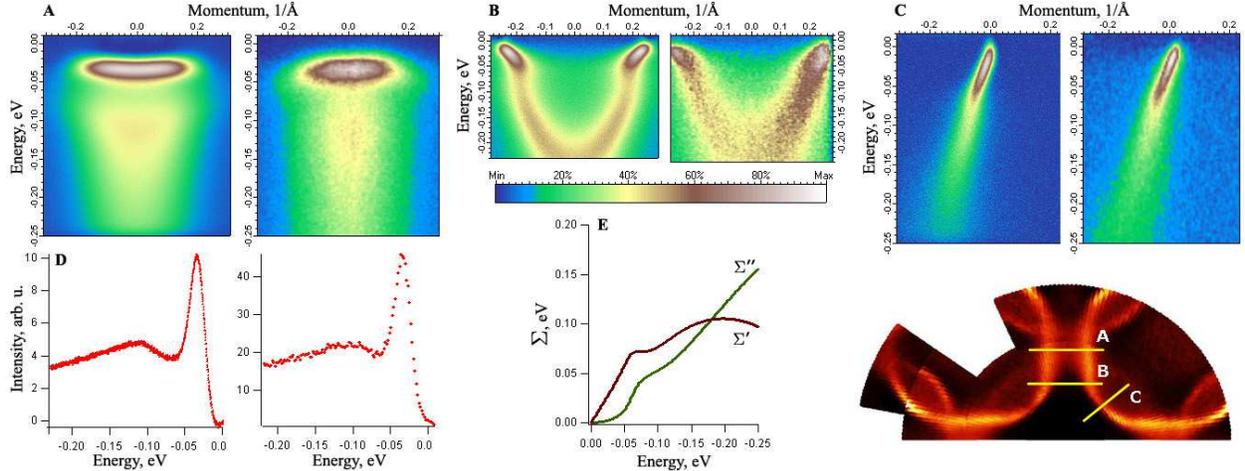}
\vspace{-5mm}\caption{The modeled (left) and experimental (right) ARPES spectra of the antibonding band at $(\pi,0)$
point (A), along a shifted cut in the same direction (B) and along the nodal direction (C) in OD Bi-2212 at 50 K. The
model images correspond to the 20 meV energy resolution and 0.025 \AA$^{-1}$ angular resolution. Panel D shows the
energy distribution curves along the dashed lines in the antinodal spectra. The Kramers-Kronig consistent real
($\Sigma'$) and imaginary ($\Sigma''$) parts of the nodal self-energy used in the calculation are shown on panel E.}
\label{f:Model}
\end{figure*}

Following these formulae, it is straightforward to conclude that knowing the real and imaginary parts of the Green's
function leads us to a comparison of ARPES results with the INS data. Although this idea was discussed earlier
\cite{AbanovChubukov99,Norman00,ManskeEremin01,NormanPepin03}, the calculations based on real ARPES data were not yet
performed. The imaginary part of the Green's function is directly related to the measured ARPES intensity (although it
can be affected by several factors). The real part can be obtained if one knows the self-energy, which is also a
routine self-consistent Kramers-Kronig procedure in the state of the art ARPES data processing
\cite{Kordyuk03,Kordyuk05}. Thus, it is possible to calculate the magnetic susceptibility response basing on the
experimental ARPES spectra.

In the superconducting state, the anomalous Green's function additionally contributes to $\chi_0$, which can not be
taken into account from ARPES data. On the other hand, as this additional term is additive, we argue that all features
observed in the first (``normal'') term, including the resonance mode, will be still present in the resulting Lindhard
function, so we content ourselves with the consideration of the first term only, even in the superconducting state.

In order to exclude the effect of matrix elements and experimental resolution, we modeled the spectra using the
self-energy based on the bare electron dispersion studied in \cite{Kordyuk05} and a self-energy model involving
quadratic scattering rate, "kink" of given width, height and position, and a density of states pile-up peak located at
the "kink" energy and characterized by the relative amplitude and width. The real part of the self-energy was
calculated by the Kramers-Kronig procedure. Self-energy parameters were specified independently for the nodal and
anti-nodal parts of the spectra, with a d-wave interpolation between these two directions. The superconducting gap was
specified in the anti-nodal direction, vanishing in the d-wave manner to the nodes. For posterior calculations, all the
free parameters were adjusted during comparison with a set of experimental ARPES spectra in order to achieve the best
correspondence (Fig.~\ref{f:Model}).

\begin{figure}[h]
\includegraphics[width=\columnwidth]{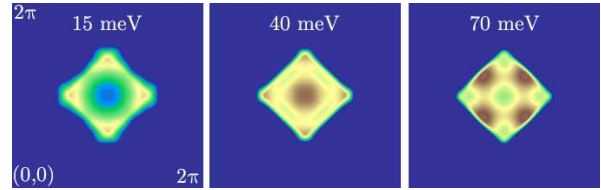}
\caption{Constant energy cuts of the dynamical spin susceptibility obtained from the renormalized Lindhard function
within the RPA approach. The center of each Brillouin zone image corresponds to the $(\pi,\pi)$ point.} \label{f:Chi}
\vspace{-2.0mm}\end{figure}

Basing on the model dataset built for optimally doped Bi-2212 at 50~K, with the superconducting gap of 30 meV,
including bonding and antibonding bands with equal intensities, we have calculated the Lindhard function in the energy
range down to 0.15~eV below the Fermi level in the whole Brillouin zone. After that we calculated the dynamical spin
susceptibility by adjusting the $J_0$ parameter trying to reproduce the resonance at $(\pi,\pi)$ and the 45$^\circ$
rotation of the incommensurate peaks that was most clearly observed in the INS experiments on YBCO
\cite{HaydenNature04}. The resulting $\chi$ (Fig.~\ref{f:Chi}) qualitatively reproduces the energy resonance at
$\sim$40~meV in the $(\pi,\pi)$ point, as well as the four peaks dispersing in the $(0,\pi)$ and $(\pi,0)$ directions
below the resonance and along the Brillouin zone diagonals for higher energies.

The~project~is~part~of~the~Forschergruppe~FOR538.

\vspace{1.5mm}

\end{document}